\documentclass[11pt,twoside]{article}

\usepackage{asp2014}

\aspSuppressVolSlug
\resetcounters

\begin{document}

\title{NASA's Long-Term Astrophysics Data Archives}

\author{L.~Rebull,$^1$ V.~Desai,$^1$ H.~Teplitz,$^1$ S.~Groom,$^1$
R.~Akeson,$^2$ G.~B.~Berriman,$^2$ 
G.~Helou,$^{3}$ D.~Imel, $^3$
J.~M.~Mazzarella,$^4$ 
A.~Accomazzi,$^5$
T.~McGlynn,$^6$
A.~Smale,$^6$ and 
R.~White$^7$
\affil{$^1$IRSA/IPAC/Caltech, Pasadena, CA, 91125, USA; \email{rebull@ipac.caltech.edu}}
\affil{$^2$NExScI/IPAC/Caltech, Pasadena, CA, 91125, USA}
\affil{$^3$IPAC/Caltech, Pasadena, CA, 91125, USA}
\affil{$^4$NED/IPAC/Caltech, Pasadena, CA, 91125, USA}
\affil{$^5$ADS/Harvard CfA, Cambridge, MA, 02138, USA}
\affil{$^6$NASA GSFC, Greenbelt, MD, 20771, USA}
\affil{$^7$MAST/STScI, Baltimore, MD, 21218, USA}}

\paperauthor{L.~Rebull}{rebull@ipac.caltech.edu}{0000-0001-6381-515X}{IRSA}{IPAC/Caltech}{Pasadena}{CA}{91125}{USA}
\paperauthor{V.~Desai}{desai@ipac.caltech.edu}{}{IRSA}{IPAC/Caltech}{Pasadena}{CA}{91125}{USA}
\paperauthor{H.~Teplitz}{hit@ipac.caltech.edu}{0000-0002-7064-5424}{IRSA}{IPAC/Caltech}{Pasadena}{CA}{91125}{USA}
\paperauthor{S.~Groom}{sgroom@ipac.caltech.edu}{}{IRSA}{IPAC/Caltech}{Pasadena}{CA}{91125}{USA}
\paperauthor{R.~Akeson}{rla@ipac.caltech.edu}{0000-0001-9674-1564}{NExScI}{IPAC/Caltech}{Pasadena}{CA}{91125}{USA}
\paperauthor{G.~B.~Berriman}{gbb@ipac.caltech.edu}{0000-0001-8388-534X}{NExScI}{IPAC/Caltech}{Pasadena}{CA}{91125}{USA}
\paperauthor{G.~Helou}{gxh@ipac.caltech.edu}{}{IPAC}{Caltech}{Pasadena}{CA}{91125}{USA}
\paperauthor{D.~Imel}{imel@ipac.caltech.edu}{0000-0002-6118-7396}{NExScI}{IPAC/Caltech}{Pasadena}{CA}{91125}{USA}
\paperauthor{J.~Mazzarella}{mazz@ipac.caltech.edu}{0000-0002-8204-8619}{NED}{IPAC/Caltech}{Pasadena}{CA}{91125}{USA}
\paperauthor{A.~Accomazzi}{aaccomazzi@cfa.harvard.edu}{0000-0002-4110-3511}{ADS}{Harvard/CfA}{Cambridge}{MA}{02138}{USA}
\paperauthor{T.~McGlynn}{tom.mcglynn@nasa.gov}{0000-0003-3973-432X}{NASA}{GSFC}{Greenbelt}{MD}{20771}{USA}
\paperauthor{A.~Smale}{alan.p.smale@nasa.gov}{}{NASA}{GSFC}{Greenbelt}{MD}{20771}{USA}
\paperauthor{R.~White}{rlw@stsci.edu}{0000-0002-9194-2807}{MAST}{STScI}{Baltimore}{MD}{21218}{USA}

\begin{abstract}
NASA regards data handling and archiving as an integral part of space
missions, and has a strong track record of serving astrophysics data
to the public, beginning with the the IRAS satellite in 1983. Archives
enable a major science return on the significant investment required
to develop a space mission. In fact, the presence and accessibility of
an archive can more than double the number of papers resulting from
the data. In order for the community to be able to use the data, they
have to be able to find the data (ease of access) and interpret the
data (ease of use). Funding of archival research (e.g., the ADAP
program) is also important not only for making scientific progress,
but also for encouraging authors to deliver data products back to the
archives to be used in future studies. NASA has also enabled a robust
system that can be maintained over the long term, through technical
innovation and careful attention to resource allocation. This article
provides a brief overview of some of NASA's major astrophysics archive
systems, including IRSA, MAST, HEASARC, KOA, NED, the Exoplanet
Archive, and ADS.
\end{abstract}

\section{Introduction}

Since at least 1983, NASA has regarded data handling and archiving as
an integral part of astrophysics space missions. This commitment now
provides the major return on the considerable investment the agency
has made over the past 20 years \citep{decadal2010}.

All astronomy archives provide sustainability. The
\citet{portals2007} concluded that a sustainable archive provides data
discovery and analysis tools; facilitates new science; contains
high-quality, reliable data; provides simple and useful tools to a
broad community; provides user support to the novice as well as to the
power user; and  adapts and evolves in response to community input.

NASA believes that an astronomy archive's job includes the following major tasks.
\begin{itemize}
\item Ingest new data, including reprocessing of old data.
\item Maintain and continue to serve a vital repository of
irreplaceable data, in which considerable investment has already been
made. This includes support for {\em observation} planning as well as,
particularly in NASA's case, {\em mission} planning. The archive must
also be a resource for original science, and a place to find high
level science products.
\item Enable cutting-edge research. NASA does this by supporting
application programming interfaces (API) and supporting the  
Virtual Observatory (VO) protocols; by providing expert user support;
by developing new and enhanced services; and by enabling multi-wavelength
projects.
\end{itemize}
The last of these is evolving quickly as the amount of data
steadily increases. The archive's mission is changing from
``search-and-retrieve'' for a user to analyze locally (on their own
machine), to doing at least some analysis {\em in situ}, in the
archive, prior to downloading.

\section{Archives Enable Science}

If one has never thought about the utilty of an archive, one might ask
whether anyone other than the proposing astronomer might be interested
in a particular data set. However, the reality is that science archives
extend the useful life of NASA's mission data indefinitely, as new
results can continue to be gleaned from the data in context with new
observations and fresh analyses.  For example, we are still learning
things from IRAS data more than 33 years after the mission ended (see
e.g.,  \citet{2015AJ....150..123R}).

\articlefigure[width=.7\textwidth]{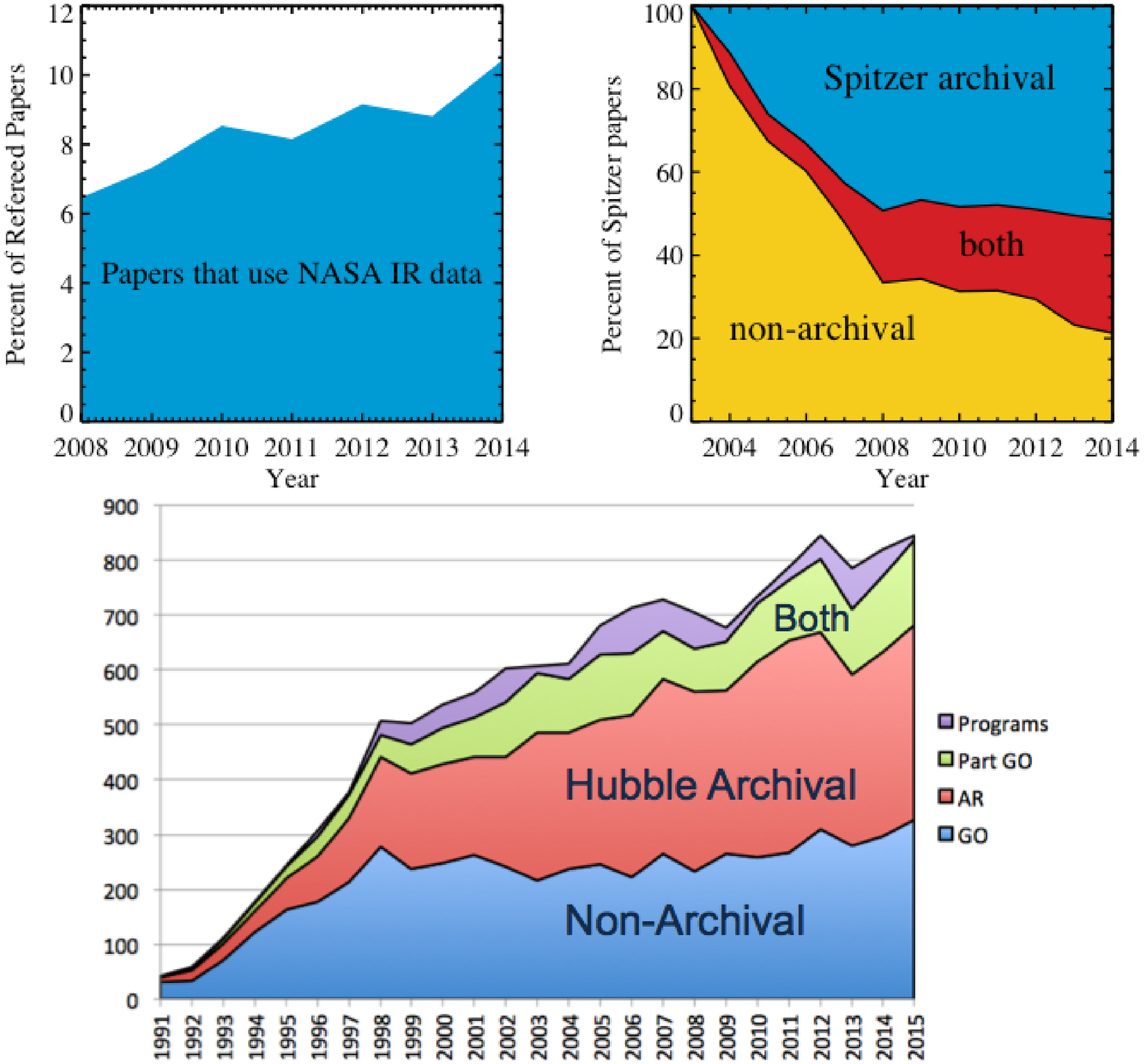}{doubledoutput}{Three plots showing
how archives double an observatory's output. Upper left: percent of
refereed journal articles as a function of time between 2008 and 2014.
As of 2014, 10\% of all refereed journal articles use data that
ultimately come from IRSA. Upper right: percent of Spitzer papers as a
function of time between 2003 and 2014. By 2008, more papers came from
archival research than PI programs. Bottom: number of Hubble papers as
a function of time between 1991 and 2015. After the first several
years of the mission, archival research dominates with more than half
the papers.}

Figure~\ref{doubledoutput} demonstrates some specific examples of how
archives {\bf double} an observatory's output. The first plot shows
the fraction of refereed astrophysics journal articles, worldwide, as
a function of time. As of 2014, 10\% of all refereed journal articles
use data that ultimately come from IRSA (this includes 2MASS, WISE,
and Spitzer). The upper right plot in Fig.~\ref{doubledoutput} shows the
fraction of Spitzer papers as a function of time. Early in the Spitzer
mission, all papers were written by program PIs; this makes sense,
since the people most equipped to process data and write papers
initially are those associated with the instrument teams and/or the
mission itself. However, by 2008, more papers came from archival
research than programs tied to specific principal investivators (PIs).
This effect is not just limited to infrared missions; the bottom panel
of Fig.~\ref{doubledoutput} shows Hubble papers as a function of time.
Once again, after the first several years of the mission, archival
research dominates, contributing more than half the papers.

Here are just a few archival science highlights:
\begin{itemize}
\item WISE and Spitzer discover the coldest brown dwarf
\citep{2014ApJ...786L..18L}
\item WISE morphological study of Wolf-Rayet nebulae
\citep{2015A&A...578A..66T} 
\item Buckyballs in a young planetary nebula using Spitzer data
\citep{2010Sci...329.1180C}
\item WISE, 2MASS, and PanSTARRS data may reveal super-void in cosmic
microwave background (CMB) cold spot seen by Planck
\citep{2015MNRAS.450..288S} 
\item The planets HR8799 b,c,d were imaged by HST in 1998; 
post-processing speckle subtraction now available provides more than an 
order of magnitude contrast improvement over what the state of the
art had been when data were taken in 1998
\citep{2011ApJ...741...55S,2015ApJ...803...31P}
\item Six years of Fermi data were combined to discover the first
extragalactic gamma-ray pulsar \citep{2015Sci...350..801F}
\end{itemize}


\section{A List of Some NASA Archives by Center}

\subsection{IPAC: IRSA}

IRSA is the NASA/IPAC Infrared Science Archive, located at IPAC at
Caltech\footnote{\url{http://irsa.ipac.caltech.edu}}. Its charter is to provide an interface to all NASA infrared
and sub-mm data sets, from $\sim$1~$\mu$m to $\sim$1~cm. It was
founded in 1993, and was the original home to IRAS data. IRSA ensures
the legacy of NASA's ``golden age'' of infrared astronomy.  IRSA
datasets are cited in about 10\% of astronomical refereed journal
articles. Through September 2016, IRSA's holdings exceed a petabyte
($>$1000 TB); there are more than 120 billion rows of catalogs.
Between January and September 2016, there have been over 33.7 million
queries, and 255 TB downloaded.

\subsection{IPAC: NED}

NED is the NASA/IPAC Extragalactic Database, located at IPAC at
Caltech\footnote{\url{http://ned.ipac.caltech.edu/ui/}}.  It is the primary hub for multi-wavelength research on
extragalactic science because it merges data from catalogs and
literature. There are thousands of extragalactic papers, with unique
measurements for millions of objects.  As of September 2016, it
contains 215 million objects with 256 million cross-identifications 
created from more than 102,000 articles and catalogs. There are 2
billion photometric data points joined into spectral energy
distributions. NED provides a thematic archive with myriad
cross-links, notes, etc., and many services tailored to extragalactic
research. Updates are released every few months.

\subsection{IPAC/NExScI: NASA Exoplanet Archive}

The NASA Exoplanet Archive is also located at IPAC at
Caltech\footnote{\url{http://exoplanetarchive.ipac.caltech.edu/}}.  It is 
focused on exoplanets and the stars they orbit, and
stars thought to harbor exoplanets. It includes Kepler data, and is
the U.S. portal to CoRoT data. It also has online tools to work with
these data, like the periodogram service. It also has a place
(Exo-FOP) for observers to upload/share data on planets and planet
candidates.

\subsection{IPAC/NExScI: KOA}

The Keck Observatory Archive (KOA) is a collaboration between NExScI
(at IPAC) and the W.\ M.\ Keck
Observatory\footnote{\url{http://exoplanetarchive.ipac.caltech.edu/}}.
It provides access to public data for all ten Keck instruments since
the Observatory saw first-light in 1994. It provides browse-quality
images of raw data, as well as browse-quality and reduced data for
HIRES, NIRC2, OSIRIS, and LWS, created by automated pipelines. An
example of contributed data is the Keck Observatory Database of
Ionized Absorption toward Quasars (KODIAQ; N. Lehner, PI). Coming
soon: NIRSPEC extracted spectra; moving target services. See poster 
\citet{P4.13_adassxxv} for more details.

\subsection{STScI: MAST}

MAST is the Mikulski Archive for Space Telescopes, and it is located
at Space Telescope Science
Institute\footnote{\url{https://archive.stsci.edu/}}. The
archive was originally established with the HST launch in 1990. It has
been multi-mission since the addition of IUE in 1998. Its mandate
includes NASA optical and UV data. It now includes Hubble, Kepler,
GALEX, IUE, FUSE, TESS, JWST, Pan-STARRS, DSS, GSC2, and more.

\subsection{GSFC: HEASARC}

HEASARC is the High Energy Astrophysics Science Archive Research
Center\footnote{\url{http://heasarc.gsfc.nasa.gov/}}. It has been
located at NASA's Goddard Space Flight Center (GSFC) since its
founding in 1990. Its mandate includes data associated with extremely
energetic cosmic phenomena ranging from black holes to the Big Bang. 
It includes data from Chandra\footnote{Note that Chandra's operations
archive is at the Chandra Science Center,
\url{http://cxc.harvard.edu/cda/}.}, XMM-Newton, Fermi, Suzaku,
NuSTAR, INTEGRAL, ROSAT, Swift, \& more than 20 others.  It merged
with Legacy Archive for Microwave Background Data Analysis (LAMBDA) in
2008; LAMBDA is the home to a variety of cosmic microwave background
radiation (CMBR) missions, including WMAP, COBE, ACT, etc.

\subsection{SAO/CfA: ADS}

ADS is the Astrophysics Data System, located at the
Harvard-Smithsonian Center for
Astrophysics\footnote{\url{http://adsabs.harvard.edu/}}.  It indexes
12 million publications in astronomy, physics, and the arXiv preprint
server. It has complete coverage of more than 100 years of astronomy
and refereed physics literature. It tracks citations, as well as
institutional and telescope bibliographies. It links to data products
at all of the other archives mentioned here, plus other non-NASA
astronomy archives around the world. It has a new interface and a new
API integrating ORCID author identifications, full-text article
searches, and analytics.

\subsection{More archives}

Other archives based at these centers, not all necessarily NASA-funded,
also follow this model.  Two examples at STScI are
Pan-STARRS (optical ground-based synoptic data) and  VLA-FIRST (radio
data). The Palomar Oschin wide-field survey is one example at IRSA; it
has three incarnations,  Zwicky Transient Facility (2017+), 
intermediate Palomar Transient Factory (iPTF; 2013-2016),  and the
Palomar Transient Factory (2009-2012).

There are, of course, many other non-NASA archives in the U.S.
(SDSS, NRAO, etc.) and around the world (CDS, ESO, ESA, etc.).
These are beyond the scope of this article, and information about
them can be found in other articles in this and prior ADASS conference
proceedings.

Also, in many cases, observers can deliver data back to these centers
for distribution, which may include data beyond original program; more
on this below.

\section{Lessons Learned}

\subsection{Easy Access and Support}

All of these archives have easy access.   Researchers at all levels
(including team members, emeriti professors, and summer students) {\em
all} need to be able to get and use data easily. Thus, these archives
need an intuitive, web-based interface, with no extra software
installation required. Users want to be able to visualize and assess
the data, using tools at the archive, but they also want to simply
download the data to their own disk as fast as possible. 

Expert help needs to be there when users need it, meaning that
documentation has to be easily found and/or helpdesk tickets promptly
answered. The archive must provide access to knowledgeable staff,
who have done science with the data products, who can (a) find
problems, and (b) pass on valuable experience to new users. Speed and
accuracy matters for the Helpdesk; people do not want to be `left
hanging.' Some questions can be very complex; acknowledging that a
person has read the ticket and is working on it is important, even if
it takes days/weeks to actually get an answer. Documentation can come
with tools and/or data releases, or in response to specific tickets
(special/unusual or frequently asked questions). Documentation should
be updated frequently in response to tickets.
Demonstrations of holdings and new tools can be provided live (at
meetings like the AAS, ADASS, DPS, etc.), or in video tutorials (IRSA
has $>$60 videos; $>$4500 views total). The complexity of science user
needs increases with time, because the `easy stuff' has already been
done -- more advanced data reduction techniques become available, or
users need to query the database in a way not envisioned when the
archive was designed.

\subsection{Data Visualization and Discovery}

Visualization at the archive is important. Science users coming to
look for original data, catalogs, or plots need to be able to find
what they are looking for. Users probably mostly come specifically to
find a particular data set, but they may also come looking for
generally anything available on their target.  They come knowing that
they need a particular item, but if data discovery is easy, they
will find more data products of use to them. Visualization helps them
assess if the data are of interest or not, before downloading (and
reading the documentation!). 

High level science products make data discovery easy and greatly
enhance the science return of the archives by making complex data sets
accessible to a wider audience of researchers.  For people who are,
for example, not experts at reducing Spitzer data, being able to find
already-reduced multi-wavelength reliable photometry of their target
means that the barriers to being a Spitzer data user are lower, and
Spitzer data get used in more projects. Hubble Legacy high-level
science products (HLSP) are used 10 times as much as more typical HST
pipeline products.  Especially when there are large, coherent projects
(such as Hubble Treasury, Spitzer Legacy, or Spitzer Exploration
Science), data reduction can be optimized for the science, and data
products become even more usable. For example, source extraction over
the entire sky has to take into account bright and faint backgrounds,
with a variety of source densities, and often depth is sacrificed for
accuracy. With a focused project, source extraction can be optimized
for that project, say, the Galactic Plane, and both depth and accuracy
is higher than with a broader project. These high level science
products products can be generated by the support center for the
telescope, or contributed by the community back to the archive. 

More recently, some delivered high level science products have
included source lists from entire missions (Spitzer, Hubble, Chandra,
Herschel, WISE, etc.). In those situations, easily combining data
across wavelengths widens the user base for each data set and deepens
the science return.

One example of such cross-wavelength advances comes from NED. In
context of assessing the completeness of the database, while studying
the fusion of extragalactic data from GALEX, SDSS, 2MASS, WISE, and
more, \citet{2016ApJ...817..109O} found super-luminous spiral
galaxies. This result was found by looking at what was already in the
archive.  

The \citet{2016ApJ...817..109O} result is just one example of two
important concepts. (1) As the data sets get bigger and bigger,
scientists won't be able to pull all of the data out of the archive to
work with it. IRAS catalogs, considered enormous at the time, can fit
on a modern iPhone without anyone particularly noticing, but WISE
catalogs top 50 billion rows. Requests to `download the entire
catalog' aren't trivial.  The mission of the archive is evolving from
a `search-and-retrieve' approach to one of `do at least some analysis
in situ.' (2) In the era of larger and larger data sets (`big data'),
there are science discoveries waiting in the archives that were never
imagined or expected by the mission or even program PIs.

\subsection{Priorities and Long-Term Commitment}

In order to set priorities, the archives rely on community feedback.
Each archive has ideas of what they would like to do next, but these
plans should be shaped by what the community needs, wants, wishes for,
or (in some cases) doesn't know they want yet. This input is collected
from mission staff members, user committees, user surveys, helpdesk
tickets, giving talks and demonstrations at conferences, and the
explicit funding review cycles, and all of that feeds into setting
priorities.

NASA as a whole (and sometimes individual missions) explicitly funds
archives, as well as archival research. The NASA ADAP (Astrophysics
Data Analysis Program) is specifically set up to fund researchers
primarily using archival data.  Having a well-designed archive \&
products can greatly enhance the research value of the dataset. In
order to achieve that goal, archives need to reduce the barriers to
usage. They need to make it easy to find the data, and make the  data
accessible  and easy to use (which speaks to reliability, units, file
format, instrumental artifacts, and documentation).  In this fashion,
NASA explicitly enables new ideas of things to do with older data.
Moreover, NASA has a strong tradition of active collaboration between
missions and archives. In practice, this means that even missions not
yet launched are thinking about optimizing the resulting science from
their future archives. 

The International Virtual Observatory Alliance (IVOA) is responsible
for standardized VO protocols for interoperability between archives
(i.e., NOT the applications that use those protocols). Tools that use
VO protocols, however, make data discovery easier. For example, users
can, with the interface they know, get access to new data elsewhere.
The VO protocols enable interoperability of tools, within archives and
across archives.  

There is also infrastructure in place to ensure that information (not
just data) flows between the people running these archives. Three
organizations that do this are the Astronomy Data Centers Executive
Committee (ADEC), the US Virtual Observatory Alliance (USVOA), and
the  NASA Astronomical Virtual Observatories (NAVO). NAVO enables
comprehensive and consistent access to all NASA data through VO
protocols. It coordinates NASA interactions with the international and
national VO communities. Figure~\ref{VOqueries} shows that the rate at
which IRSA receives VO-protocol queries is increasing with time.

\articlefigure[width=.7\textwidth]{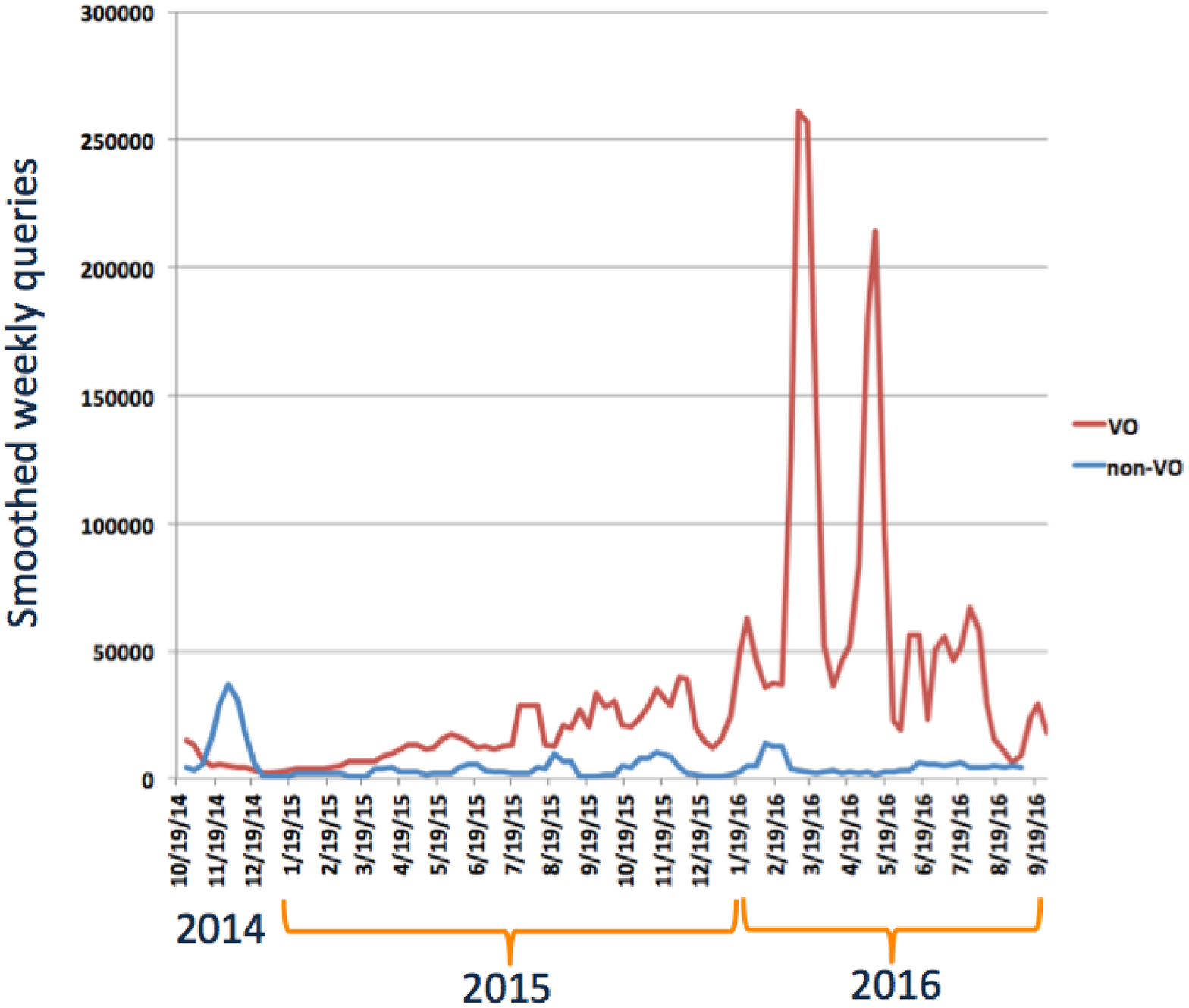}{VOqueries}{The
smoothed weekly queries at IRSA as a function of time for VO queries
(red) and non-VO queries (blue). There are increasing numbers of VO
queries.}

\subsection{Keeping it Running}

As archives, and the science they enable, grow more sophisticated,
there are more and more API queries made of the archives. APIs, or
application program interfaces, to the archives enable programs,
scripts, or even users at the command line to query the archives. 
Scripted access to archive data enables complex projects. However, it
also enables rapid queries. New users of the API, in particular, can
launch inadvertent denial of service attacks on the servers.  A real
life example (from IRSA) is more than 70,000 requests over 10 hours,
or an average of 2 per second. The servers must be watched to ensure
that no one user is robbing the rest of the community of bandwidth and
computing resources.  Empirically, IRSA has many users making small
requests via the APIs, but a few users make enormous requests.

The archive must be kept running 24/7 while improving it, or
`assembling the plane while it is in flight.' The archives aim to
increase the audience and their usage of the archive while keeping it
within existing resources. As a result, the archives must be efficient
in how they use resources. One example of how IRSA does this is to use
the same software across multiple data sets (see talk by X. Wu).

IRSA has an interactive user interface, re-using pieces developed by
other projects at IPAC. NED is experimenting with machine learning;
NED needs to absorb data that is embedded in free-form text, where the
tables are not standard. For example, the position of the object may be
given in a column whose header is one of RA, Ra, ra, R.A., or
something else entirely. NED has a pilot project to apply machine
learning to classify data and facilitate their extraction. The
archives must improve scalability, extensibility, and data
prospecting, and they must accomplish greater integration of
functionality and content across systems. For example, ADS, as an
archive focused on the literature, bridges the ``data'' world of
astronomy archives and the ``publishing'' world of scholarly literature.
As such it integrates content and functionality relevant to both.
ORCID offers an example of a standard which has been promoted by
publishers to help with author disambiguation; integration of the
claiming and indexing in ADS means that the astronomy community has an
``easy'' way to create the claims using a trusted platform. Searching
by object name using SIMBAD TAP offers an example of integration of
cross-archive functionality in the new ADS using VO standards. ADS
also provides embedding of publisher images via APIs. 

The archives have to be ready to ingest new data from the community.
At IRSA, Spitzer Legacy programs changed the astronomy culture by
mandating that products be delivered back to the community. Now such
data deliveries are a common feature of Spitzer proposals. As
discussed above, these deliveries bring these data to a larger
audience via the central Spitzer archive. But,  IRSA has to have
resources to ingest these products. The expense is not necessarily in
hardware resources but in the time it takes to educate the people
delivering the products. The delivery has to be well-organized and
documented, not just for the people at IRSA operationally ingesting
the data, but for all future users of the data. For people who
frequently make deliveries, this is (now) easy.  It is not necessarily
easy for people new at making deliveries. IRSA has developed tools to
help people learn, but it still takes time and often hand-holding.
Complexity is not just about size. And an unanticipated side effect is
that you can get optical and UV data out of the Spitzer archive
(SINGS, LVL).

\subsection{What's Next: Big Data}

The era of ``big data'' is here for some missions, and certainly it
has arrived when one considers collectively the data across all of the
NASA astrophysics archives. IRSA has already invested in data
visualization services to help people identify and experiment with
data quickly, and decide whether they want to download the data or
not. Planning for big data includes identifying the most
critical needs of users, including increased analysis at the archive
facilitated by user workspaces, and richer services for in-situ
analysis.  All of the archives are thinking about this in some way.

\section{Summary}

Long-term, sustainable archives greatly increase the return on
observatory investment, doubling the science return in published
papers. Having robust, reliable support for both expert and novice
users pays off. User support by instrument experts is crucial for the
sucessful use of the data/archives by the wider astronomy community.
Standardization of tools within an archive increases efficiency.
Interoperability between archives increases access to data sets and
facilitates multi-mission analysis. High level data
products can expand the reach of large data sets. The archives are
seeing a shift in approach from `search and retrieve' to `analyze in
situ.'

\bibliography{I3.1}  

\begin{thebibliography}{}
\expandafter\ifx\csname natexlab\endcsname\relax\def\natexlab#1{#1}\fi
\expandafter\ifx\csname url\endcsname\relax
  \def\url#1{\texttt{#1}}\fi
\expandafter\ifx\csname urlprefix\endcsname\relax\def\urlprefix{URL }\fi
\providecommand{\eprint}[2][]{\url{#2}}

\bibitem[{{Cami} et~al.(2010){Cami}, {Bernard-Salas}, {Peeters}, \&
  {Malek}}]{2010Sci...329.1180C}
{Cami}, J., {Bernard-Salas}, J., {Peeters}, E., \& {Malek}, S.~E. 2010,
  Science, 329, 1180

\bibitem[{{Committee for a Decadal Survey of Astronomy and
  Astrophysics}(2010)}]{decadal2010}
{Committee for a Decadal Survey of Astronomy and Astrophysics} 2010
  (Washington, DC: National Academies Press)

\bibitem[{{Committee on NASA Astronomy Science Centers}(2007)}]{portals2007}
{Committee on NASA Astronomy Science Centers} 2007 (Washington, DC: National
  Academies Press)

\bibitem[{{Fermi LAT Collaboration} et~al.(2015){Fermi LAT Collaboration},
  {Ackermann}, {Albert}, {Baldini}, {Ballet}, {Barbiellini}
  et~al.}]{2015Sci...350..801F}
{Fermi LAT Collaboration}, {Ackermann}, M., {Albert}, A., {Baldini}, L.,
  {Ballet}, J., {Barbiellini}, G., et~al. 2015, Science, 350, 801

\bibitem[{{Luhman}(2014)}]{2014ApJ...786L..18L}
{Luhman}, K.~L. 2014, \apjl, 786, L18. \eprint{1404.6501}

\bibitem[{{Ogle} et~al.(2016){Ogle}, {Lanz}, {Nader}, \&
  {Helou}}]{2016ApJ...817..109O}
{Ogle}, P.~M., {Lanz}, L., {Nader}, C., \& {Helou}, G. 2016, \apj, 817, 109.
  \eprint{1511.00659}

\bibitem[{{Pueyo} et~al.(2015){Pueyo}, {Soummer}, {Hoffmann}, {Oppenheimer},
  {Graham}, {Zimmerman}, {Zhai}, {Wallace}, {Vescelus}, {Veicht}, {Vasisht},
  {Truong}, {Sivaramakrishnan}, {Shao}, {Roberts}, {Roberts}, {Rice}, {Parry},
  {Nilsson}, {Lockhart}, {Ligon}, {King}, {Hinkley}, {Hillenbrand}, {Hale},
  {Dekany}, {Crepp}, {Cady}, {Burruss}, {Brenner}, {Beichman}, \&
  {Baranec}}]{2015ApJ...803...31P}
{Pueyo}, L., {Soummer}, R., {Hoffmann}, J., {Oppenheimer}, R., {Graham}, J.~R.,
  {Zimmerman}, N., {Zhai}, C., {Wallace}, J.~K., {Vescelus}, F., {Veicht}, A.,
  {Vasisht}, G., {Truong}, T., {Sivaramakrishnan}, A., {Shao}, M., {Roberts},
  L.~C., Jr., {Roberts}, J.~E., {Rice}, E., {Parry}, I.~R., {Nilsson}, R.,
  {Lockhart}, T., {Ligon}, E.~R., {King}, D., {Hinkley}, S., {Hillenbrand}, L.,
  {Hale}, D., {Dekany}, R., {Crepp}, J.~R., {Cady}, E., {Burruss}, R.,
  {Brenner}, D., {Beichman}, C., \& {Baranec}, C. 2015, \apj, 803, 31.
  \eprint{1409.6388}

\bibitem[{{Rebull} et~al.(2015){Rebull}, {Carlberg}, {Gibbs}, {Deeb}, {Larsen},
  {Black}, {Altepeter}, {Bucksbee}, {Cashen}, {Clarke}, {Datta}, {Hodgson}, \&
  {Lince}}]{2015AJ....150..123R}
{Rebull}, L.~M., {Carlberg}, J.~K., {Gibbs}, J.~C., {Deeb}, J.~E., {Larsen},
  E., {Black}, D.~V., {Altepeter}, S., {Bucksbee}, E., {Cashen}, S., {Clarke},
  M., {Datta}, A., {Hodgson}, E., \& {Lince}, M. 2015, \aj, 150, 123.
  \eprint{1507.00708}

\bibitem[{{Rizzi}(2017)}]{P4.13_adassxxv}
{Rizzi}, L. 2017, in ADASS XXVI, edited by TBD (San Francisco: ASP), vol. TBD
  of ASP Conf. Ser., TBD

\bibitem[{{Soummer} et~al.(2011){Soummer}, {Brendan Hagan}, {Pueyo},
  {Thormann}, {Rajan}, \& {Marois}}]{2011ApJ...741...55S}
{Soummer}, R., {Brendan Hagan}, J., {Pueyo}, L., {Thormann}, A., {Rajan}, A.,
  \& {Marois}, C. 2011, \apj, 741, 55. \eprint{1110.1382}

\bibitem[{{Szapudi} et~al.(2015){Szapudi}, {Kov{\'a}cs}, {Granett}, {Frei},
  {Silk}, {Burgett}, {Cole}, {Draper}, {Farrow}, {Kaiser}, {Magnier},
  {Metcalfe}, {Morgan}, {Price}, {Tonry}, \& {Wainscoat}}]{2015MNRAS.450..288S}
{Szapudi}, I., {Kov{\'a}cs}, A., {Granett}, B.~R., {Frei}, Z., {Silk}, J.,
  {Burgett}, W., {Cole}, S., {Draper}, P.~W., {Farrow}, D.~J., {Kaiser}, N.,
  {Magnier}, E.~A., {Metcalfe}, N., {Morgan}, J.~S., {Price}, P., {Tonry}, J.,
  \& {Wainscoat}, R. 2015, \mnras, 450, 288. \eprint{1405.1566}

\bibitem[{{Toal{\'a}} et~al.(2015){Toal{\'a}}, {Guerrero}, {Ramos-Larios}, \&
  {Guzm{\'a}n}}]{2015A&A...578A..66T}
{Toal{\'a}}, J.~A., {Guerrero}, M.~A., {Ramos-Larios}, G., \& {Guzm{\'a}n}, V.
  2015, \aap, 578, A66. \eprint{1503.06878}

\end{thebibliography}

\end{document}